\documentclass[conference]{IEEEtran}
\IEEEoverridecommandlockouts
\usepackage{cite}
\usepackage{amsmath,amssymb,amsfonts}
\usepackage{algorithmic}
\usepackage{graphicx}
\usepackage{textcomp}
\usepackage{xcolor}
\usepackage{pgfplots}
\usepackage{subfigure}
\usepackage{listings}
\usepackage{subfigure}

\usepackage{textcomp}
\usepackage{xcolor}
\usepackage{fancyhdr}
\usepackage{multirow}
\usepackage{textcomp}
\usepackage{multirow}
\usepackage{caption,tabularx,booktabs}
\usepackage{pifont}

\usepackage{algorithmic}
\usepackage{algorithm}
\usepackage{amsmath}
\usepackage{pgfplots}
\usepackage{subfigure}
\usepackage{url}

\pgfplotsset{width=7cm,compat=1.8}
\usetikzlibrary{pgfplots.groupplots}

\def\BibTeX{{\rm B\kern-.05em{\sc i\kern-.025em b}\kern-.08em
    T\kern-.1667em\lower.7ex\hbox{E}\kern-.125emX}}
\begin{document}

\title{Making Your Program Oblivious: a Comparative Study for Side-channel-safe Confidential Computing}

\author{\IEEEauthorblockN {A K M Mubashwir Alam}
	\IEEEauthorblockA{\textit{Computer Science} \\
		\textit{Marquette University}\\
		Milwaukee, WI, USA \\
		mubashwir.alam@marquette.edu}
	\and
	\IEEEauthorblockN  {Keke Chen}
	\IEEEauthorblockA{\textit{Computer Science} \\
		\textit{Marquette University}\\
		Milwaukee, WI, USA \\
		keke.chen@marquette.edu}
}

\maketitle

\begin{abstract}
	Trusted Execution Environments (TEEs) are gradually adopted by major cloud providers, offering a practical option of \emph{confidential computing} for users who don't fully trust public clouds. TEEs use CPU-enabled hardware features to eliminate direct breaches from compromised operating systems or hypervisors. However, recent studies have shown that side-channel attacks are still effective on TEEs. An appealing solution is to convert applications to be \emph{data oblivious} to deter many side-channel attacks. While a few research prototypes on TEEs have adopted specific data oblivious operations, the general conversion approaches have never been thoroughly compared against and tested on benchmark TEE applications. These limitations make it difficult for researchers and practitioners to choose and adopt a suitable data oblivious approach for their applications. To address these issues, we conduct a comprehensive analysis of several representative conversion approaches and implement benchmark TEE applications with them. We also perform an extensive empirical study to provide insights into their performance and ease of use.
\end{abstract}


\section{Introduction}

Confidential computing enables users to enjoy public clouds without the need to trust cloud providers' security infrastructure. Researchers are actively developing cryptographic approaches to secure processing in untrusted platforms, such as Homomorphic Encryption  \cite{brakerski11} and Secure Multiparty Computation (SMC) \cite{huang11,mohassel17}. While recent cryptographic methods, e.g., hybrid protocols \cite{niko13sp,mohassel17,sharma19}, are getting more efficient, pure software-based solutions are still too expensive to be practical for complex computational tasks or data-intensive applications \cite{sharma21survey}. 


More recently, Trusted Execution Environment (TEE) \cite{sgx-explained} emerges as a more practical solution for confidential computing. TEE utilizes CPUs' new hardware features to securely isolate a user's application from the cloud system. Therefore, even if an entire system, including the operating system or hypervisor, is compromised, the adversary cannot access the application. Major CPU manufacturers have implemented the TEE concept in their recent CPUs, e.g., Intel SGX and AMD SEV. Correspondingly, TEE-enabled servers are increasingly available in major public clouds, e.g., Azure provides SGX-enabled servers, and Google adopts AMD SEV.

While TEE performs much more efficiently than software-based cryptographic approaches, recent studies have also identified several side-channel attacks \cite{xu15channel,shinde16,bulck17,bulck17step,chen18}. Although the \emph{TEE enclave} cannot be directly breached, side channels are still there -- The enclave interacts with untrusted memories and file systems, and the CPU cache is still shared among processes/virtual machines owned by different users. Thus, attackers can utilize such controlled channel attacks, e.g., manipulating page faults and page-table entries and exploiting the flaws of modern CPU's micro-architecture execution optimization. Powerful attacks like Foreshadow \cite{bulck18} and Load Value Injection \cite{vanbulck2020lvi} can combine memory/cache footprints and CPU speculative execution to extract the secrets in TEE execution.

So far, countermeasures on side-channel attacks are limited to specific applications \cite{ahmad18,ohrimenko15} or firmware fixes at the micro-architectural \cite{vanbulck2020lvi}. Among the candidate solutions, data-oblivious algorithms and applications appear attractive and promising. Regular programs' data flow and execution paths vary according to input data, i.e., a specific input value may trigger different steps to execute. In contrast, data-oblivious algorithms' data flow and execution paths are invariant to the input. This data obliviousness property can potentially help address many side-channel issues, as we will discuss in Section \ref{sec:side-channel}. 

Nevertheless, it is challenging for users to develop an oblivious solution for the following reasons. First, it's unclear how complex to compose an oblivious data program manually. Although recent TEE-related studies \cite{ohrimenko16, sasy18, constable18} have indicated some oblivious primitives that one can use to compose an oblivious solution, the complexity and the efforts to develop such a solution are unclear. Second, automated approaches can help convert regular programs to oblivious ones, but it's unclear how practical they are for TEE applications. Third, the quality of automatically generated oblivious solutions is also a concern. Oblivious programs generally cost more than their non-oblivious equivalent in terms of performance and memory. Low-quality conversion may also result in a higher performance penalty. There is no systematic study to answer these questions for TEE-based applications.

\textbf{Contributions.} We conduct a comprehensive analysis and empirical study to compare several data oblivious solutions for TEE applications. The result will help researchers and practitioners understand the benefits and limitations of current solutions, possibly identify new research topics and assess the strategies for adopting the side-channel-safe TEE solutions. 

Specifically, we first analyze whether and how data obliviousness can address side-channel attacks. Then, we summarize four representative solutions: the manual composition approach, the compiler approach, the circuit approach, and the application-framework approach and their characteristics in terms of performance, ease of use, and maturity for application.

Finally, we develop an evaluation benchmark that includes basic oblivious operations, compute-intensive tasks, and data-intensive tasks. Then, we apply different oblivious program conversion approaches to the benchmark and evaluate the resulting oblivious programs' performance and ease of use. Our study reveals the strengths and weaknesses of different oblivious solutions and provides guidelines for selecting suitable techniques under different scenarios. 

In the remaining sections, we will first present the background knowledge for our approach (Section \ref{sec:background}), then dive
in the details of how data oblivious solutions help to protect side-channels (Section \ref{sec:data-oblivious-for-side-channel}), then discuss different oblivious approaches (Section \ref{sec:oblivious-approaches}), and, finally, perform the experimental evalution (Section \ref{sec:experimental-eval}) and conclusion (Section \ref{sec:conclusion}).

\section{Preliminaries} \label{sec:background}
%

This section presents the necessary preliminaries for understanding the paper. We will give the related background knowledge before analyzing oblivious solutions. In the following, we will introduce the concept of Trusted Execution Environments (TEE), the status of TEE development and deployment, and the effect of side-channel attacks in TEEs.

\subsection{Trusted Execution Environment}  
Trusted Execution Environment (TEE) is a hardware-based solution for executing code in a secure environment where powerful adversaries cannot access code or data within this secure area. Using TEEs, a user can run their sensitive computations in a TEE called Enclave, which uses a hardware-assisted mechanism to preserve the privacy and integrity of enclave memory. With TEEs, users can pass encrypted data into the Enclave, decrypt it, compute with plaintext data, encrypt the result, and return it to the untrusted cloud components. TEEs isolate private reserved memory for secure applications from other system components, such as operating systems and hypervisors. When the operating system or other system applications want to access the dedicated private memory, the CPU restricts the access and redirects to some abort memory page. Therefore, TEE applications can perform plaintext calculations without compromising privacy and security. 

However, without verifying the correctness of the cloud hardware and the user binary, the remote user still cannot trust the TEE. The Remote Attestation procedure establishes the trust between the TEE hardware and the user. Using remote attestation, the user can verify if the cloud provider is using certified TEE hardware and if the program running in an enclave is from a digitally signed binary. During remote attestation, a secret key is generated by Diffie-Hellman key exchange between the Enclave and the remote user to establish a secure channel for the follow-up communication between the user and the Enclave. 

Major cloud platforms have provided different types of TEE-enabled servers. Intel SGX is one of the popular TEE implementations. Since 2015, SGX has been available in most Intel CPUs. Unlike Intel SGX, the other two major TEEs, such as ARM TrustZone and AMD PSP, rely on secure hardware and a secure operating system. As AMD integrated ARM TrustZone as an extension of CPU \cite{amd2020} and later renamed it Platform Security Processor (PSP), the underlying technology of both systems remains similar. While all TEE implementations feature complete memory isolations from the system components and remote attestation to establish trust, they still suffer from side-channel attacks.

\subsection{Side Channel Attacks on TEEs} \label{sec:side-channel}
Since the advent of TEEs, many studies have explored the weaknesses of TEE side channels. While passive adversaries can exploit some attacks \cite{cash15,zheng17,olga15} by only observing interactions between TEEs and other system components, the assumption of TEEs enables more powerful attacks to be performed, some of which can even retrieve plaintext information directly from the Enclave. Based on the attack strategies, these attacks can be categorized as (i) memory/cache-targeted attacks and (ii) microarchitecture-level attacks. In memory/cache-targeted attacks, the attacker exploits the interactions between TEEs and untrusted memory or applications and observes enclave memory page loading and CPU cache usages. Microarchitecture-level attacks utilize modern CPU features, such as CPU transient memory execution \cite{bulck18}, to retrieve fine-grained information from the low-level cache lines. We will discuss more details in the next section. 

\section{Data Oblivious Solutions for Side Channel Protection}  \label{sec:data-oblivious-for-side-channel}

\subsection{Threat Model}
\begin{figure}[t]
	\centerline{\includegraphics[width=.7\linewidth]{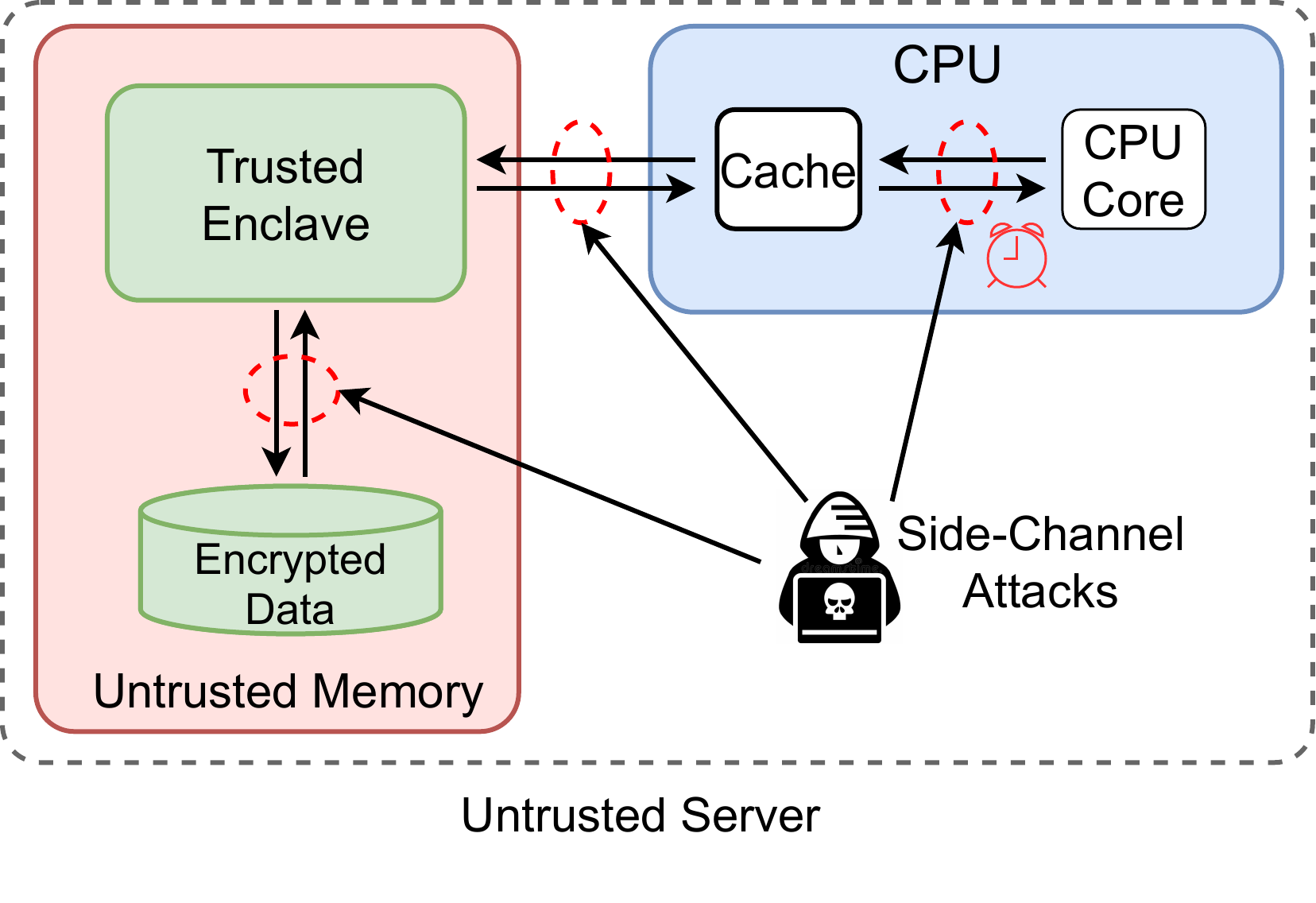}}
	\caption{TEE, side channels, and the threat model. 
	}
	\label{fig:sgx_side_channels}
\end{figure}

Users may run confidential computation tasks in an untrusted cloud server, where the server's OS or hypervisor can be compromised. The goal is to preserve data and program's integrity and confidentiality while availability is out of concern. A typical TEE, such as Intel SGX, provides a hardware-protected memory area, i.e., the \emph{enclave} \cite{sgx-explained}, and guarantees the integrity of the data and computation running inside the enclave. While adversaries cannot directly access the enclave, they can still glean information via side channels, such as memory access patterns and CPU caches. However, cache-based attacks target all CPUs (regardless of having TEEs or not) and thus need manufacturers' micro-architecture level fixes. In contrast, the exposure of memory access patterns is inevitable as enclaves have to interact with the untrusted memory area. It's also reasonable to assume that attackers cannot access the cloud server physically, e.g., attaching a device to the server or touching the motherboard, which excludes all attacks based on physical accesses. Figure \ref{fig:sgx_side_channels} illustrates the threat model. 

\subsection{Data Obliviousness}

\textbf{Definition.} The execution path and data flow of a (data) oblivious program do not change with different input data and parameter settings. When all the steps of an algorithm or mechanism do not depend on input data, one cannot determine the nature of the data by observing the steps of that algorithm. Thus, oblivious solutions can effectively protect from attacks depending on data-dependent access patterns.

\textbf{Oblivious Primitives. } The goal of developing data oblivious programs is to eliminate any data-dependent operations. We list the primitives that data oblivious programs heavily depend on. 

\begin{itemize}
	\item \textbf{Address-based Access}. This operation includes array element access or data block access. Exposing the position of accessed data is the fundamental access pattern. A naive solution is to iterate over the whole data structure to hide the actual accessed position. In contrast, Oblivious RAM (ORAM) \cite{oded96} has been a well-accepted primitive for more efficiently hiding accessed addresses. It can effectively reduce the cost of oblivious access to $O(\log N)$ for a structure of $N$ data items. ORAM has been used in a few TEE-based solutions to hide access patterns.  
	\item \textbf{Data-dependent Branching.} Most programs contain  data-dependent branching statements. Depending on the different inputs, a program execution may choose different paths, resulting in distinct access patterns. The following code snippet shows how an attacker can utilize the branching access pattern. 
	
\begin{lstlisting}[frame=none, numbers=none]
if (a >= b){
  // swap a and b, and 
  // the page access can be observed.
}else{
  // no page access.
}
\end{lstlisting}

	The common method uses the CPU's conditional move (CMOV) instructions to eliminate the branching statements. A simplified example is shown as follows:
	
\begin{lstlisting}[frame=none, numbers=none]
//if (a < b) x = a else x = b
CMOVL x, a
CMOVGE x, b
\end{lstlisting}

	A few studies \cite{olga16,rane15,alam21} have used CMOV instructions  to provide code-level obliviousness for branching statements. Without specific conditional jumps, CMOV instructions move the source operand to the destination when a conditional flag is set. However, regardless the flag is set or not, it reads the source operand. Therefore, the access to the source operand cannot be used to infer whether the source is copied to the destination or not. Ohrimenko et al. \cite{olga16} also designed library functions \emph{omove} and \emph{ogreater} to wrap up the CMOV instructions for conveniently converting the branching statements. Notably, a completely oblivious branching execution needs to run both branches and select the desired result with the above method, which often leads to very high costs. 
	
	\item \textbf{Circuit.} Circuits are considered a natural way to hide access patterns, as the circuit execution activates the gates in a certain order regardless of the input values \cite{heath22}. The branching statement is readily implemented with a bitwise multiplexer. However, oblivious memory access imposes significant challenges. Many solutions implement linear scan so far \cite{simon18,circ22,buscher18}, which incurs very high costs.

	\item \textbf{Oblivious algorithms.} Task-specific oblivious algorithms are methods specifically designed to work with a specific task or a data structure. They work more efficiently than solutions composed of general primitives such as ORAM. For example, MergeSort can be converted to be oblivious by simply replacing every memory interaction of the merge phase with ORAM and unwinding data-dependent loops with fixed iteration loops. However, this direct conversion can be much more expensive than a specially designed oblivious sorting algorithm, such as BitonicSort \cite{batcher68}. Similarly, frequently used data-intensive operations, such as \emph{join} and \emph{group by}, can have more efficient dedicated oblivious versions. 
	
\end{itemize}

\subsection{Can Data Obliviousness Address TEE Side-channel Attacks?}

\textbf{Against Memory Targeted Attacks.} Memory-targeted attacks glean and utilize access patterns within the system memory. TEE applications store data in an encrypted form outside the TEE. When encrypted messages are accessed from memory, i.e., between TEE and untrusted memory and even within TEE, an adversary who controls the operating system can observe the data access patterns and possibly extract sensitive information by manipulating page-fault interrupts \cite{cash15,shinde16}. Page-table entries \cite{bulck17}.  For distributed data-intensive applications,  Ohrimenko et al. \cite{olga15} also demonstrated how sensitive information, such as age group, birthplace, and marital status, can be extracted from MapReduce programs by only observing the network flow and memory skew. 

These attacks all depend on the differential access patterns observed via the side channels. For example, an important step in KMeans clustering is to find the nearest centroid and update the temporal cluster information for each training data. A straightforward way of accessing cluster information creates data-dependent branching based on cluster ids. If each cluster object resides in separate memory pages, an attacker can exploit cluster id-dependent branches by observing memory page access patterns. Thus, the adversary can estimate the cluster size, which may be sensitive to the user. However, a simplified oblivious version of this step hides the secret dependent branch by accessing each cluster object with a CMOV operation. Thus, the attacker cannot distinguish which cluster-id is being updated for the training data.

\textbf{Against Cache Attacks}. Cache-based side-channel attacks \cite{nilsson20} had been long exploited before TEE became popular. The basic mechanism of cache attacks remains the same for systems with or without TEE. The main idea of the cache attack is to load the system memory into the CPU cache and perform a time analysis by loading different byte values to retrieve the value of the previously loaded memory, such as \emph{Prime+Probe} \cite{fangfei15} and \emph{Flush+Reload} \cite{yarom14} methods. 

Like regular applications, TEE is also vulnerable to cache-based side-channel attacks. Since the last level cache (LLC) is a shared resource, an attacker can exploit fine-grained information at a specific stage of the program by probing the data access time in each cache line. However, in a data-oblivious algorithm, all the steps and data accesses are fixed. Thus, an attacker cannot distinguish the secret value and dummy access from the cache-level timing at a given time. For example, a cache attack cannot distinguish the secret-dependent block IDs or block data from dummy ones if accessed through oblivious RAM \cite{sasy18,ahmad18}.

\textbf{Against Micro-architectural Attacks.}
Some powerful attacks exploit the CPU's micro-architecture to retrieve secrets from TEE applications. Foreshadow \cite{bulck18} exploits meltdown-type \cite{lipp18} attacks on TEE applications. Load Value Injection (LVI) \cite{vanbulck2020lvi} is the most recent attack on Intel SGX that successfully retrieves the secrets from the victim's Enclave within the victim's address space. The CPU's micro-architectural buffer must be prepared with some attacker-controlled secret value to perform the LVI attack. These attacks are powerful enough to extract plain text information from the TEE without physical access. Manufacturers have issued microarchitectural-level firmware patches for some  \cite{vanbulck2020lvi,bulck18} of these attacks. 

However, not all micro-architectural attacks can be prevented from firmware-level patches. Some micro-architectural-level attacks still utilize access patterns. For example, Bulck et al. \cite{van18} show that, by exploiting the timing of micro-architectural instructions, attackers can observe secret-dependent branches at the CPU instruction level. This type of micro-architectural-level attacks cannot succeed if the application developer hides the data-dependent branches with oblivious solutions. Therefore, oblivious programs can still help mitigate these attacks.

\section{Making Your Program Oblivious }  \label{sec:oblivious-approaches}

While the fundamentals of data-oblivious operations are clear, developing a practical solution is challenging for several reasons. First, it requires the developer to have basic knowledge of every data-dependent part of their programs and the risk of leaking a particular access pattern. Second, converting the program to an oblivious one can be complex and error-prone. We investigate the existing candidate approaches and summarize the following four most representative ones for developing oblivious solutions: (i) manual composition (or manual approach), (ii) compiler approach, (iii) circuit approach, and (iv) framework approach.

\subsection{Manual Composition}
In manual composition, developers need to learn all the knowledge of sensitive access patterns and the methods of converting them to be oblivious. These approaches may vary depending on the applications' related access pattern problems. The key challenges of this approach are to manually analyze the access pattern problem for every line of the code and replace the vulnerable parts with their oblivious alternatives. Developers may also need to experiment with different oblivious primitives to determine the most efficient one. It's also necessary to verify whether the conversion is complete with a tool such as ObliCheck \cite{oblicheck}. 

Several problem-specific manual compositions have been reported to address the access-pattern based attacks on TEE applications. To protect the random access over block data in untrusted memory, the developer can implement Oblivious RAM that works with TEE, e.g., ZeroTrace \cite{sasy18,ahmad18}, that hides which block is read or written by shuffling memory blocks during each access. Other problem-specific algorithms have also been used to hide access patterns of specific tasks, such as Oblivious Sorting \cite{batcher68}, Oblivious Filter \cite{zheng17}, Oblivious Join \cite{Krastnikov20}, etc. CMOV-based oblivious branching is also actively applied in developing specific machine algorithms \cite{ohrimenko16} and addressing the in-enclave access patterns \cite{sasy18,alam21}, with wrapped functions such as oblivious move, greater, swap, etc. 

Manually applying oblivious solutions enables the designing of both memory and performance-efficient application. However, manually analyzing sensitive data access and code vulnerabilities and applying oblivious solution is time-consuming, domain-expertise demanding, and sometimes error prone. Developers may utilize existing oblivious libraries \cite{constable18} to reduce manual efforts.

\subsection{Compiler Approach}

Compiler techniques (e.g., static code analysis) can be used to minimize manual efforts. The current compiler approaches for generating oblivious code follow two directions: automate the manual composition approach, and hide code and data access patterns via randomization.

The first category of approaches includes Raccoon\cite{rane15} and Ghostrider \cite{liu15} for automating the manual process. They utilize static code analysis to detect vulnerabilities, i.e., the primitive operations that need to be obfuscated, and then apply suitable oblivious measures, as we have discussed in the manual approach. They also use a few methods to optimize performance. For example, they often provide an option to allow developers to annotate the parts to be obfuscated; based on the array size, the compiler can decide to use linear scan or ORAM to hide the accessed element.  

Another compiler approach is to randomize access patterns and program execution paths \cite{orenbach19, ahmad19, brasser2017dr} by using primitives like ORAMs. For example, Obfuscuro \cite{ahmad19} divides the memory and code pages into two classes and achieves obliviousness via ORAM. Similarly, Dr. SGX \cite{brasser2017dr} and CoSMIX \cite{orenbach19} also use data randomization techniques to achieve fully automated memory obfuscation. The non-deterministic run-time access patterns successfully hinder the adversary from extracting any information. 

These approaches are mostly experimental, not fully achieving the design goal yet. First, while the compiler approach avoids most manual efforts, it may not generate the most efficient oblivious solution. In particular, the randomization approach often results in significantly high costs. Second, since the compilers depend on static analysis to identify the sensitive code blocks, they may apply unnecessary obfuscation due to a lack of context awareness and the limitation of static analysis tools. Son et al.\cite{oblicheck} showed how static analysis-based methods failed to recognize standard oblivious algorithms and were marked as non-oblivious, leading to unnecessary obfuscation in the mitigation phase.

The ultimate goal of the compiler approach is to entirely free developers from complex and expensive manual efforts, which is appealing. Unfortunately, existing compilers have not reached the goal yet. In fact, we could not find a stable open-source implementation for any of the mentioned approaches. Nevertheless, we still believe this is a promising direction.

\subsection{Circuit Approach} 
In many cryptographic approaches \cite{huang11,mohassel17}, circuits have been used as a building block for secure evaluation. Boolean circuits are naturally oblivious as they execute all the paths \cite{simon18,circ22}. Since automated program-to-circuit conversion tools have been built for the cryptographic approaches, such as ObliVM \cite{liu15}, CircC \cite{circ22}, and HyCC \cite{buscher18}, we wonder whether this approach can also be a candidate for TEE-based applications. 

Cryptographic circuit compilers ensure many things, including variable mutations, conditional branching, loops, loops with breaks, early returns, and random array access, are oblivious. We briefly describe some of the key features:
\begin{itemize}
	\item Variable Mutations. In a regular program, the value of a variable can be updated; however, in Boolean circuits, values are mutation free. When required to update a variable, it uses versions of the variable. For example, if the previous value of x was zero and then assigned to ten, it will be converted to $x_0 = 0$, $x_1 = 10$.   
	\item Branching. To protect the branching attack, researchers implement guarded execution, executing both branches and selecting the result based on condition. 
	\item Loops. In circuits, each gate is executed only once. Thus, circuits do not have any loops. Compilers unroll the loops to linear execution for bounded iterations. If you do not provide a bound, it will unroll the loop up to the size of data types. For example, a byte-type loop variable will unroll for 256 iterations.
	\item Loops with breaks. Loops with breaks are common in programming. While older compilers do not support break statements, modern compilers support break/continue commands. In short, compilers use a similar approach that resolves conditional branching. Expressly, compilers turn the loop into a breakable block. When there is a breakable block, it will execute all iterations, except it will use guarded execution to hide which value will be chosen last.
	\item Random Array access. That array access depends on the input. It requires O(n) operations. Since the value is not constant, array access is implemented using an n-sized multiplexer with the index as the multiplexer selector.
\end{itemize}

Cryptographic circuit compilers such as CircC \cite{circ22} and HyCC \cite{buscher18} can convert regular C programs to executable C circuit programs. However, very few TEE-related studies \cite{selo20,felsen19} have applied circuits as an access pattern protection mechanism. We show in experiments that the performance can be a significant concern due to three reasons (1) the extensive uses of the above methods to ensure obliviousness; (2) the size of the generated circuit is large, proportional to the input data size; (3) executing a circuit in a software mode is inherently slow. 

\subsection{Framework Approach}
For TEE-based data-intensive applications, the framework approach can be a valid candidate \cite{schuster15,ohrimenko16,dinh15,zheng17,alam21}. We refer ``framework'' here to the well-known big data processing frameworks, such as MapReduce \cite{dean04} and Spark \cite{zaharia10}. We have witnessed two types of framework approaches. 

The first type aims to extend the big data frameworks to take advantage of TEEs, which will have to handle side-channel attacks, as well. VC3 \cite{schuster15} applied this strategy for modifying the Hadoop \cite{white09} system. M2R \cite{dinh15} targets the problem of access-pattern leakage in the shuffling phase of VC3 and proposes to use the oblivious schemes for shuffling. Another significant work, Opaque \cite{zheng17}, tries to revise Spark for SGX. They focus on the data access patterns between computing nodes, illustrate how adversaries can use these to infer sensitive information in the encrypted data and design data-oblivious methods to address these attacks. 
These frameworks have a shared weakness: only a few data processing components of the framework were moved to the TEE, leaving most parts of the framework in untrusted areas under serious threats. However, re-implementing the frameworks with the data obliviousness guarantee is too expensive to be practical.

The second type of framework approach utilizes a data-intensive framework to simplify the development of data oblivious solutions. Specifically, the framework serves as the middleware to hide the complexity of data-oblivious processing. The developer only needs to spend much less effort implementing data-oblivious TEE applications. SGX-MR \cite{alam21} utilizes the MapReduce processing model to regulate application dataflow so that the framework can integrate the application-independent data-oblivious protection mechanisms in the framework code to protect any data mining algorithms that can be cast to the MapReduce framework. Developers only need to handle a much smaller number of and often simpler access patterns in the application-specific map and reduce functions. 

This framework approach has a few unique benefits: (1) It significantly reduces the complexity of the manual approach for data-intensive processing; (2) It can integrate both the latest big-data processing techniques and data-oblivious solutions at the framework level, which is transparent to developers; (3) It can incorporate data-intensive optimization techniques in the framework implementation, e.g., most block-based operations. Thus it can be more efficient than directly applying fully automated compiler or circuit approaches that are unaware of data-intensive features.

We use Table \ref{tab:summary} to summarize the four approaches qualitatively in terms of ease of use, performance, and unique features. In experiments, we will see more quantitative results.

\begin{table}[h!]
	\begin{center}
		\centering
		\caption{Summary of existing data oblivious solutions.
		}
		\label{tab:summary}
		\begin{tabular}{|l|l|l|l|l|l|}
			\hline
			\textbf{Method} & \textbf{Easy use}  & \textbf{Performance} & \textbf{Other features}\\
			\hline
			Manual & Low  & High & Require  expertise, \\ &&& time consuming \\
			\hline
			Compiler & High & Mid & May incur unnecessary \\ &&& obfuscation \\
			\hline			
			Circuit & Mid & Low & compilation very slow, \\ &&& circuit size greater than data \\
			\hline
			F/W & High & High & only domain specific \\ &&& functionalities \\
			\hline
		\end{tabular}
	\end{center}
\end{table}

\section{Experimental Evaluations}  \label{sec:experimental-eval}
The experimental evaluation has the following goals.
\begin{itemize}
	\item Observe how different conversion approaches perform on the primitive operations, which are the building blocks of an oblivious program. These operations include (i) oblivious data access, (ii) oblivious branching, and (iii) basic oblivious algorithms, such as sorting. The primitive operations will be discussed under compute-intensive and data-intensive workloads. 
	\item Understand how the oblivious programs generated with different approaches perform. We evaluate the performance with compute-intensive and data-intensive benchmark applications.
	\item Investigate the ease of use, i.e., the developers' efforts, via multiple measures, e.g., the line of code (LOC), the number of sensitive code blocks susceptible to access pattern leakages, and LOC overhead to achieve mitigation techniques.
\end{itemize}

\subsection{Experiment Setup}
The experiments were conducted on a Linux machine with an Intel(R) Core(TM) i7-8700K CPU of a 3.70GHz processor and 16 GB of DRAM. The TEE environment is Intel SGX v1.0, and the Linux version is Ubuntu 22.04. 

\textbf{Approach Implementation.} The \emph{unprotected} approach is the simple implementation without considering obliviousness. The \emph{manual} composition approach has used the following primitive operations: linear scan and ORAM for random array accesses, the omove/ogreater functions \cite{ohrimenko16,sasy18,alam21} for oblivious branching, and BitonicSort for sorting, to convert the implementation. We used ZeroTrace's open-source implementation for ORAM on SGX, and the HyCC circuit generator \cite{buscher18} to convert a plain implementation to a circuit for the circuit-based evaluation. For data-intensive applications, we adopted the SGX-MR framework in the evaluation. Unfortunately, we could not find a working open-source implementation for the published compiler approaches. Due to the sheer complexity of implementing such a compiler, we have not included the compiler approach in the evaluation. 

\begin{itemize}
	\item \emph{Oblivious Operations.} We evaluate the three basic oblivious operations: random array access (block access in data-intensive workloads), branching, and sorting. This evaluation should help understand the performance of each approach at low-level operations. 
	
	\item \emph{Sample Applications.} In addition to the basic oblivious operations, we also include four applications for compute-intensive and data-intensive workloads, two for each category respectively. Each application will be a certain blend of primitive oblivious operations. (1) The compute-intensive applications include the Edit Distance computation for varying lengths of strings, which has a complexity of $O(N^2)$ for the string length $N$, and the All-Pair Shortest Path algorithm, Floyd-Warshall, with a complexity of $O(N^3)$ for $N$ nodes in the graph. (2) The data-intensive applications include the WordCount and  KMeans clustering algorithms. 
\end{itemize}

In the following, we will organize the results in terms of compute-intensive and data-intensive workloads. The primitive operations are discussed under each category correspondingly.

\subsection{Results for Compute-Intensive Workloads}

Compute-intensive workloads use a relatively small dataset that can be fit into the TEE memory. We choose three basic operations and then two well-known applications to evaluate the performance of different oblivious methods. 

\textbf{Random Array Access.} Accessing array elements is one of the basic operations of most applications. Figure \ref{fig:obli-array-access} shows the execution time of various oblivious implementation methods. First, the expenses of all oblivious solutions are significantly higher than the unprotected version. ORAM shows performance advantages over linear scans for larger sizes. For smaller sizes, ORAM's maintenance cost becomes relatively high, resulting in higher overall costs than linear scans.  The circuit version of array access also uses linear scans. However, executing a circuit at the software level is clearly very inefficient regardless of the array size.

\newcolumntype{C}[1]{>{\centering\arraybackslash}p{#1}}
		
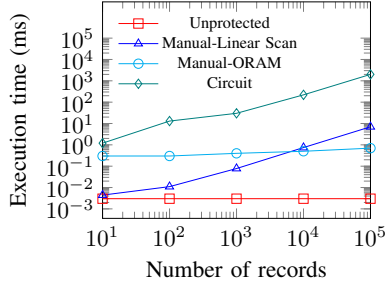
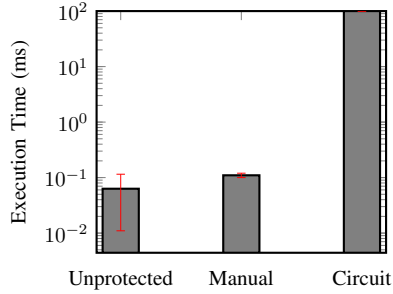
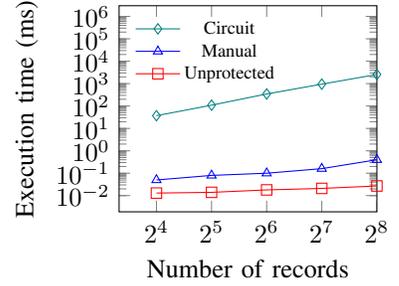
\begin{figure*}[t]
	\centering
	\begin{tabular}{C{.31\textwidth}C{.3\textwidth}C{.31\textwidth}}
		
		\subfigure[Oblivious array access. Record size 8 bytes.]{
			\resizebox{0.3\textwidth}{!} {
				
					\begin{tikzpicture}
					\begin{axis}[
						width=\linewidth,
						title style={align=center},
						scaled ticks=false,
						xlabel={Number of records},
						ylabel={Execution time (ms)},
						xmin=10, xmax=100000,
						ymin=0, ymax=5000000,
						xtick={0, 10, 100, 1000, 10000, 100000, 1000000, 10000000},
						ytick={0, 0.001, 0.01, 0.1, 1, 10, 100, 1000,  10000, 100000 },
						legend style={ draw = none, fill=none},
						legend style={nodes={scale=0.7, transform shape}},	
						legend pos=north west,
						legend columns=1,
						xmode=log,
						ymode=log
						]

						\addplot[
						color=red,
						mark= square,
						]
						coordinates {
							(10, 0.003)
							(100, 0.003)
							(1000, 0.003)
							(10000, 0.003)
							(100000, .003)
						};
						\addlegendentry{Unprotected}

						\addplot[
						color=blue,
						mark= triangle,
						]
						coordinates {
							(10, 0.00441)
							(100, 0.01101)
							(1000, 0.07755)
							(10000, 0.74556)
							(100000, 7)
						};
						\addlegendentry{Manual-Linear Scan}
						
						\addplot[
						color=cyan,
						mark= o,
						]
						coordinates {
							(10, 0.3)
							(100, 0.3)
							(1000, 0.4)
							(10000, 0.5)
							(100000, 0.7)
						};
						\addlegendentry{Manual-ORAM}
						
						\addplot[
						color=teal,
						mark= diamond,
						]
						coordinates {
							(10, 1.2)
							(100, 13)
							(1000, 30)
							(10000, 220)
							(100000, 2000)
						};
						\addlegendentry{Circuit}
						
					\end{axis}
				\end{tikzpicture}
				\label{fig:obli-array-access}
			}
		}&
		
			
			\subfigure[Oblivious branching.]{
				\resizebox{0.3\textwidth}{!} {
		\begin{tikzpicture} 
		\begin{semilogyaxis} [
			width=\linewidth,
			xticklabel style={text height=2.5ex},
			xtick= {0, 1, 2, 3},
			xticklabels = {not, Unprotected, Manual, Circuit},
			ylabel={Execution Time (ms)},
			bar width=.6cm,
			x = 2cm,
			y = .4cm,
			ymin = 0,
			ymax = 100,
			line width=1pt,
			log origin=infty
			]
			
			\addplot [
			draw=black, fill=gray,
			ybar,
			error bars/.cd,
			y dir=both,
			y explicit,
			error bar style={red},
			] coordinates {
				(1,0.063) +- (0.052,0.052) 
				(2, 0.11) +- (0.01,0.01) 
				(3, 100) +- (0.01,0.01)
			};
		\end{semilogyaxis}
		\end{tikzpicture}
		\label{fig:obli-code}
				}
			}&
			
			\subfigure[Oblivious sorting. Record size 8 bytes.]{
				\resizebox{0.3\textwidth}{!} {
				\begin{tikzpicture}
				\begin{axis}[
					width=(\linewidth* .9),
					title style={align=center},
					scaled ticks=false,
					xlabel={Number of records},
					ylabel={Execution time (ms)},
					xmin=10, xmax=256,
					ymin=0, ymax=3000000,
					xtick={0, 16, 32, 64, 128, 256},
					ytick={0, 0.01, 0.1, 1, 10, 100, 1000, 10000, 100000, 1000000 },
					legend style={ draw = none, fill=none},
					legend style={nodes={scale=0.7, transform shape}},	
					legend pos=north west,
					legend columns=1,
					xmode = log,
					ymode=log,
					 log basis x={2}
					]

					\addplot[
					color=teal,
				mark= diamond,
					]
					coordinates {
						(16, 37)
						(32, 109)
						(64, 345)
						(128,946)
						(256,2563)
					};
					
					\addlegendentry{Circuit}

					\addplot[
					color=blue,
				mark= triangle,
					]
					coordinates {
						(16, 0.05)
						(32, 0.08)
						(64, 0.1)
						(128, 0.16)
						(256, 0.4)
					};
					\addlegendentry{Manual}
					
					\addplot[
					color=red,
				mark= square,
					]
					coordinates {
						(16, 0.013)
						(32, 0.014)
						(64, 0.018)
						(128, 0.021)
						(256, 0.027)
					};
					\addlegendentry{Unprotected}
					
				\end{axis}
			\end{tikzpicture}
			\label{fig:obli_sort}
			}}
		\end{tabular}
		\caption{Performance evaluation of in-memory core operations.}
	\end{figure*}

\textbf{Branching.} We design a benchmark program to evaluate the impact of oblivious branching. In a loop, a branching statement decides to execute one of the two functions: one with a low cost and the other with a high cost. We repeated the experiments a few times. Naturally, the unprotected version's performance varies over different runs. Figure \ref{fig:obli-code} shows that the manual approach, which uses CMOV instructions, is relatively efficient. However, the circuit approach shows multiple orders of magnitude higher costs.

\textbf{Sorting.} The manual approach adopts BitonicSort. Since the circuit approach can convert any algorithm to its oblivious version, we have considered different sorting algorithms for the circuit approach. It turns out BitonicSort is also the most efficient one in circuit form. Figure \ref{fig:obli_sort} shows the manual approach is much faster than the circuit approach on BitonicSort.

\textbf{Edit Distance.} Edit distance uses dynamic programming to compute the distance between two sequences, whose complexity is $O(N^2)$ for sequences of length $N$. It's a typical high-complexity algorithm working with a relatively small amount of memory. In Figure \ref{fig:edit_distnace}, we notice that the manual method is close to the unprotected version, but again the circuit approach is much more expensive than others.

\newcolumntype{C}[1]{>{\centering\arraybackslash}p{#1}}
\begin{figure*}[t]
	\centering
	\begin{tabular}{C{.4\textwidth}C{.4\textwidth}}
		
		\subfigure[Edit distance.]{
			\resizebox{0.4\textwidth}{!} {
				\begin{tikzpicture}
				\begin{axis}[
					width=(\linewidth* .9),
					title style={align=center},
					scaled ticks=false,
					xlabel={Sequence lengths},
					ylabel={Execution time (ms)},
					xmin=10, xmax=100,
					ymin=0, ymax=1000000,
					xtick={10, 20,  40, 60, 80, 100},
					ytick={0, 0.01, 0.1, 1, 10, 100, 1000, 10000, 100000, 1000000 },
					legend style={ draw = none, fill = none},
						legend style={nodes={scale=0.9, transform shape}},	
					legend pos=north west,
					legend columns=1,
					ymode=log
					]
					
					\addplot[
					color=red,
				mark= square,
					]
					coordinates {
						(10,0.1)
						(20,0.28)
						(30,0.40)
						(40,0.6)
						(50,0.73)
						(60,.91)
						(100,1.6)
					};
					\addlegendentry{Unprotected}
					
					\addplot[
					color=blue,
				mark= triangle,
					]
					coordinates {
						(10, 0.3)
						(20, 0.6)
						(30, 0.8)
						(40, 1)
						(50, 1.3)
						(60, 1.8)
						(100, 3.7)
					};
					\addlegendentry{Manual}
					
					\addplot[
				color=teal,
				mark= diamond,
					]
					coordinates {
						(10,30)
						(20,310)
						(30,962)
						(40,2000)	
						(50,3000)
						(60,4320)
						(100, 8500)
					};
					\addlegendentry{Circuit}
					
				\end{axis}
			\end{tikzpicture}
			\label{fig:edit_distnace}
			}
		}&
			
			\subfigure[Floyd-Warshall all-pair shortest path.]{
				\resizebox{0.4\textwidth}{!} {
						\begin{tikzpicture}
						\begin{axis}[
							width=(\linewidth)*0.9,
							title style={align=center},
							scaled ticks=false,
							xlabel={Number of Nodes},
							ylabel={Execution time (ms)},
							xmin=-10, xmax=400,
							ymin=0, ymax=10000,
							xtick={0, 50, 100, 200, 300, 400, 500},
							ytick={0, 0.01, 0.1, 1, 10, 100, 1000, 10000 },
							legend style={ draw = none},
							legend style={nodes={scale=0.9, transform shape}},	
							legend pos=south east,
							ymode=log
							]
							
							\addplot[
							color=red,
							mark= square,
							]
							coordinates {
								(5, 0.015)
								(10, 0.02)
								(15, 0.033)
								(20, 0.057)
								(25, 0.093)
								(30, 0.14)
								(100,4)
								(200,29)
								(300,94)
								(400,218)
								(500,417)
							};
							\addlegendentry{Unprotected}
							
							\addplot[
							color=blue,
							mark= triangle,
							]
							coordinates {
								
								(5,0.05)
								(10,0.21)
								(15,0.44)
								(20,1)
								(25,2)
								(30,3.3)
								(60, 26)
								(100,120)
								(200,966)
								((300,3261)
								(400,7728)
								(500,15248)
								
							};
							\addlegendentry{Manual}
							
							\addplot[
						color=teal,
						mark= diamond,
							]
							coordinates {
								(5,77)
								(10,755)
								(15,2643.74)
								(20,6895)
							};
							
							\addlegendentry{Circuit}
							
						\end{axis}
					\end{tikzpicture}
					\label{fig:warshall}
			}}
		\end{tabular}
		\caption{Performance evaluation of compute-intensive applications.}
	\end{figure*}
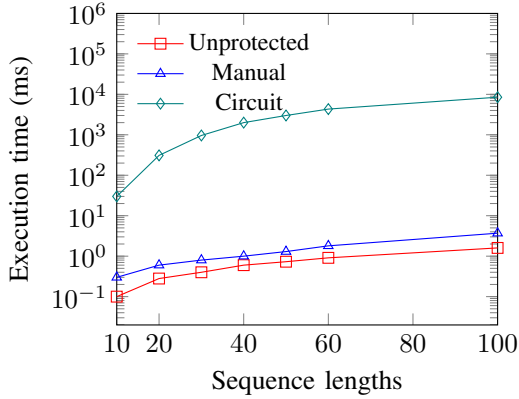
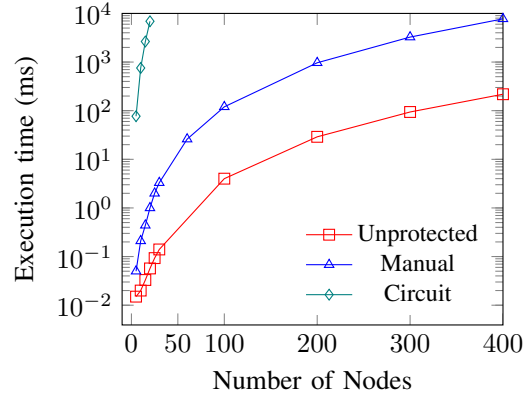

\textbf{All-pair shortest path.} The all-pair shortest path Floyd-Warshall algorithm is expensive with a complexity of $O(N^3)$ for $N$ nodes. Figure \ref{fig:warshall} shows the manual approach is significantly slower than the unprotected one, but at a manageable scale. In contrast, the circuit approach is too expensive to handle larger $N$.

\subsection{Results for Data-intensive Workloads}
For data-intensive workloads, we evaluate two basic operations: block-level random access and block-based external sorting. In application-based evaluations, we also include the framework approach: SGX-MR. 

\textbf{Random block access.} Similar to the evaluation on array access, we include linear scan and ORAM methods for the manual approach. Figure \ref{fig:random-block} clearly shows the manual-ORAM approach performs much better. However, due to the large data size, the gap between oblivious approaches and the unprotected is also large. 

\newcolumntype{C}[1]{>{\centering\arraybackslash}p{#1}}
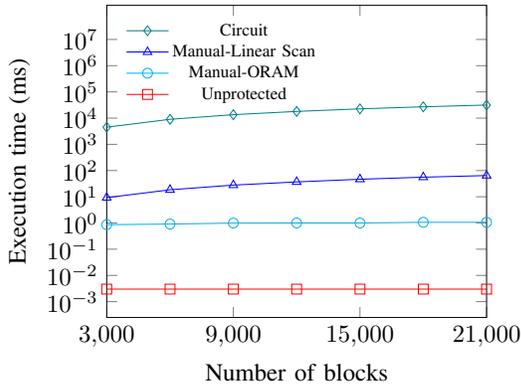
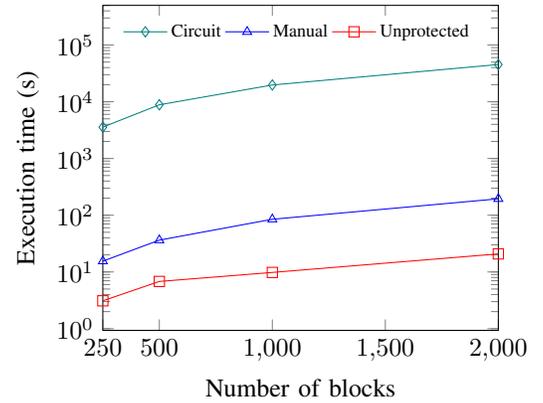
\begin{figure*}[t]
	\centering
	\begin{tabular}{C{.5\textwidth}C{.5\textwidth}}
		
		\subfigure[Comparing random Access over blocks. Block size 1KB.]{
			\resizebox{0.4\textwidth}{!} {
					\begin{tikzpicture}
					\begin{axis}[
						scaled ticks=false,
						xlabel={Number of blocks},
						ylabel style={align=center},
						ylabel={Execution time (ms)},
						xmin=3000, xmax=21000,
						ymin=0, ymax=200000000 ,
						xtick={3000,9000, 15000, 21000},
						ytick={10000000, 1000000, 100000,10000, 1000, 100, 10, 1, .1, .01, .001},
						legend style={nodes={scale=0.7, transform shape}},
						legend style={ draw = none, fill=none},
						legend pos=north west,
						legend columns = 1,
						ymode=log,
						]

						\addplot[
						color=teal,
						mark= diamond,,
						]
						coordinates {
							(3000,4500)
							(6000,9000)
							(9000,13500)
							(12000,18000)
							(15000,22500)
							(18000,27000)
							(21000,31500)
						};
						\addlegendentry{Circuit}
						
						\addplot[
							color=blue,
							mark= triangle,
						]
						coordinates {
							(3000,9.26)
							(6000,18.45)
							(9000,27.84)
							(12000,36.8)
							(15000,46.54)
							(18000,55.82)
							(21000,64.44)
						};
						\addlegendentry{Manual-Linear Scan}
						
						\addplot[
							color=cyan,
							mark= o,
						]
						coordinates {
							(3000,.86)
							(6000,.91)
							(9000,.99)
							(12000,1)
							(15000,1)
							(18000,1.06)
							(21000,1.06)
						};
						\addlegendentry{Manual-ORAM}
						
						\addplot[
							color=red,
							mark= square,
						]
						coordinates {
							(3000,.003)
							(6000,.003)
							(9000,.003)
							(12000,.003)
							(15000,.003)
							(18000,.003)
							(21000,.003)
						};
						\addlegendentry{Unprotected}
						
					\end{axis}
				\end{tikzpicture}
				\label{fig:random-block}
			}
		}&
		
		\subfigure[Block-based oblivious sorting. Block size 1KB with 75 words (records) per block.]{
			\resizebox{0.4\textwidth}{!} {
					\begin{tikzpicture}
					\begin{axis}[
						title style={align=center},
						scaled ticks=false,
						legend style={nodes={scale=0.7, transform shape}},
						xlabel={Number of blocks},
						ylabel={Execution time (s)},
						xmin=250, xmax=2000,
						ymin=0, ymax=500000,
						xtick={250, 500, 1000, 1500, 2000},
						ytick={1, 10, 100, 1000, 10000, 100000, 1000000},
						legend style={ draw = none},
						legend pos=north west,
						legend columns=3,
						ymode = log
						]
						\addplot[
						color=teal,
						mark=diamond,
						]
						coordinates {
							(250, 3570)
							(500,	8840)
							(1000,	19740)
							(2000,	45300)
						};
						\addlegendentry{Circuit}
						\addplot[
						color=blue,
						mark=triangle,
						]
						coordinates {
							(250, 15.5)
							(500,	36.37)
							(1000,	84.83)
							(2000,	194.63)
						};
						\addlegendentry{Manual}    
						\addplot[
							color=red,
							mark= square,
						]
						coordinates {
							(250, 3.1)
							(500, 6.8)
							(1000, 9.8)
							(2000, 20.8)
						};
						\addlegendentry{Unprotected}
					\end{axis}
				\end{tikzpicture}
				\label{fig:block-sort}
		}}
	\end{tabular}
	\caption{Performance evaluation of core operations for data-intensive applications.}
\end{figure*}

\textbf{Block-based External Sorting.} Next, we implement block-level sorting to understand the sorting cost in block-level operations. We used 1-KB blocks filled with string data. The manual approach adopts a block-level BitonicSort algorithm with oblivious in-block operations \cite{alam21}, while the circuit approach converts the block-level BitonicSort algorithm. In Figure \ref{fig:block-sort}, we observe that the circuit approach is still orders of magnitude higher than the manual approach, while the manual approach is about ten times slower than the unprotected one. 

\textbf{WordCount.} For application-level evaluation, we take a fixed amount of input, 500 1KB blocks, each of which is filled with random text. As the MapReduce-based solution is the most efficient one for the WordCount problem, the manual approach essentially duplicates the processing encoded in the framework of SGX-MR, which uses BitonicSort for the intermediate sorting of word-count pairs. As a result, the manual approach has an almost identical cost to the SGX-MR approach. Certainly, SGX-MR significantly simplifies the developer's coding efforts. Again, the circuit approach is too expensive to be a practical solution.

\newcolumntype{C}[1]{>{\centering\arraybackslash}p{#1}}
\begin{figure*}[t]
	\centering
	\begin{tabular}{C{.4\textwidth}C{.4\textwidth}}
		
		\subfigure[ Application-level perfromance for Wordcount. Number of blocks 500 with 75 words/block.]{
			\resizebox{0.3\textwidth}{!} {
					\begin{tikzpicture} 
					\begin{axis} [
						width=\linewidth,
						xticklabel style={text height=3ex},
						xtick= {1, 2, 3, 4},
						xticklabels = {{Manual}, {SGX-MR}, {Circuit}},
						xticklabel style={align=center},
						ylabel={Execution Time (s)},
						bar width=.6cm,
						x = 2cm,
						ymode = log,
						]
						
						\addplot [
						draw=black, fill=gray,
						ybar
						] coordinates {
							(1, 16)
							(2, 15)
							(3, 9872)
						};
					\end{axis}
				\end{tikzpicture}
				\label{fig:wordcount}
			}
		}&
		
		\subfigure[Application-level performance for KMeans. 4000 1KB-Blocks with eight bytes per record, and five clusters.]{
			\resizebox{0.3\textwidth}{!} {
				\begin{tikzpicture} 
					\begin{axis} [
						width=\linewidth,
						xticklabel style={text height=3ex},
						xtick= {1, 2, 3, 4},
						xticklabels = {{Manual}, {SGX-MR}, {Circuit}},
						xticklabel style={align=center},
						ylabel={Execution Time (s)},
						bar width=.6cm,
						x = 2cm,
						ymin = 1,
						ymode = log,
						]
						
						\addplot [
						draw=black, fill=gray,
						ybar
						] coordinates {
							(1, 2) 
							(2, 9.2)
							(3, 85740)
						};
					\end{axis}
				\end{tikzpicture}		
				\label{fig:kmeans}
		}}
	\end{tabular}
	\caption{Performance evaluation of data-intensive applications.}
\end{figure*}
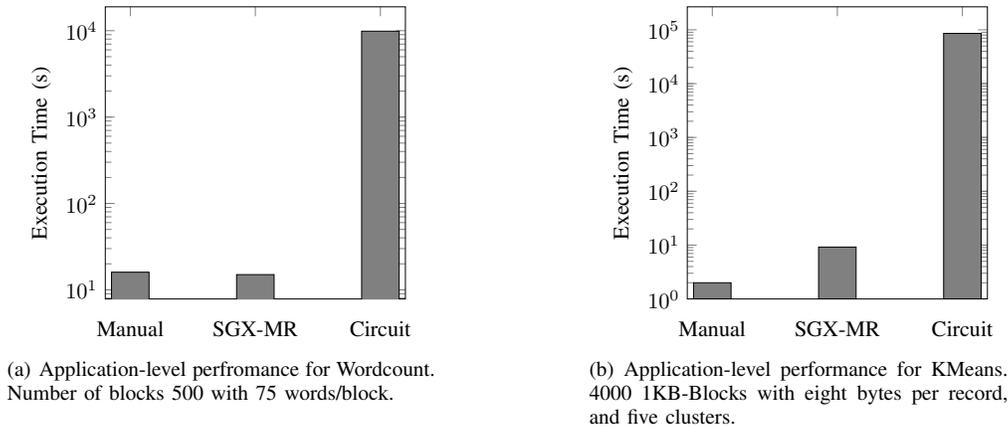

\textbf{KMeans.} We use 4000 1k data blocks consisting of $34\times10^4$ records and five clusters. Due to the small number of clusters, we use hash and ORAM for aggregation (check Appendix \ref{code:oram-hash-algo}), which appears more efficient than sorting-based aggregation in SGX-MR.  Figure \ref{fig:kmeans} shows this manual approach performs best among the candidate techniques.

\subsection{Developers' effort to achieve data oblivious solutions}

We are also curious about how easy a developer can use each of these approaches. This evaluation does not include the extra time learning the different approaches -- apparently, developers need to take a significant amount of time to learn the manual approach and the framework approach.

Instead, we look at the result of developing the evaluated applications to understand the difficulty levels of using different approaches. We also assume developers will use a library of oblivious primitives, e.g., ORAM, oblivious branching, and oblivious sorting. The use of library will  also significantly reduce the line of code (LOC) for the manual and framework approaches.

	\begin{table*}[h] 
	\begin{center}
	\caption{Summary of developers' effort to implement the oblivious solutions. The table represents the total line of code as LOC and the number of access-pattern sensitive segment in the code as AP. LOC-overhead means the lines used to achieve data obliviousness.}

		\label{tab:effort-cost}
		\begin{tabular}{|c|c|c|c|c|c|c|c|c|c|}
			\hline
			\textbf{Application} & \multicolumn{3}{c|}{\textbf{Manual}} & \multicolumn{3}{c}{\textbf{Circuit}} & \multicolumn{3}{|c|}{\textbf{Framework}} \\
			& LOC & LOC-Overhead & AP & LOC & LOC-Overhead & AP & LOC & LOC-Overhead & AP  \\
			\hline
			Edit Distance &58 & 28 & 4 & 48 & 0 & - & - & - & - \\ 
			\hline
			All-Pair Shortest Path & 47 & 15 & 1 & 36 & 0 & - & - & - & - \\ 
			\hline
			Word Count &277 & 21 & 6 & 155 & 0 & - & 22 & 0 & 0 \\ 
			\hline
			KMeans  & 330 & 24 & 4 & 263 & 0 & - & 58 & 6 & 1 \\ 
			\hline
			
		\end{tabular}
	\end{center}
\end{table*}

Table \ref{tab:effort-cost} summarizes the additional effort a developer need to achieve data oblivious applications. In the following we will discuss the compared data oblivious strategies. 

\begin{itemize}
	
	\item \textbf{Manual Composition.} Manual composition requires domain knowledge on TEE side channels. The table shows the manual approach requires identifying one to six sensitive code segments and hiding the access pattern with data-oblivious alternatives. This approach requires the developer to write more lines of code than other approaches, even when the oblivious library is used. 
	
	\item \textbf{Circuit.} The circuit approach is fully automated, and the developer does not need to do any additional work. 
	
	\item \textbf{Framework.} With a framework like SGX-MR, the developer only focuses on small pieces of application-specific code, such as the map and reduce functions, dramatically reducing the developer's burden compared to the manual approach. The framework software contains fully optimized oblivious code that is transparent to developers and shared by all SGX-MR applications. Table \ref{tab:effort-cost} shows by using SGX-MR for framework-level protection, the developer does not require any LOC overhead for word count, and for KMeans developer only needs to add six lines of code to solve one access pattern issue.
\end{itemize}

Overall, the circuit approach is the easiest to use as it does not require any additional effort from the developer. The manual approach involves a lot of efforts in analyzing the original code and conducting the conversion. In contrast, the framework approach hides many details with the framework implementation and minimizes the developer's efforts. However, it does require the developer to learn to use the framework first.

\section{Conclusion}  \label{sec:conclusion}
Data oblivious programs provide excellent defenses against several side-channel attacks targeting TEE applications. However, developing oblivious programs is challenging. We have analyzed four representative approaches that can help developers convert non-oblivious programs to oblivious ones. Among these approaches, we consider performance and ease of use (and possibly readiness to use) are critical measures. Our experimental results show that: (1) The manual composition approach gives the best performance guarantee, while developers must fully understand the access pattern of every part of their code and learn the corresponding conversion method; (2) The framework approach for data-intensive applications achieves a good balance between performance and ease of use; (3) The circuit approach is theoretically sound, but extremely expensive in practice; and (4) the compiler approach is promising, but not mature enough for practical use. We hope our analysis and evaluation will help both practitioners to decide their solutions and researchers to explore potential issues.

We consider a few promising research directions. (1) The compiler approach is the most appealing one, as it aims to make the conversion process fully transparent to developers and the converted program to have a good performance close to the manual composition approach. (2) Another direction is data oblivious libraries/frameworks and software tools to automate the conversion process as possible and minimize the developers' manual efforts. The framework approach is an excellent example of this direction.

%

\bibliographystyle{abbrv}
\bibliography{./paper, ./sgx_sc}

\appendices
\section{Sample ORAM-Hash algorithms}
\label{code:oram-hash-algo}	

	\begin{algorithm}
		\caption{Buffer Management for ORAM-Hash algorithms}\label{alg:oram_buffer}
		\label{code:buffer-management}
		\begin{algorithmic}[1]
			\STATE Buffer contains a working block $B$ for new records, and a cache of $m$ blocks $C$.
			\STATE \textbf{Function} $\mathrm{GetBlock} (block\_id)$  
			\IF{ $block\_id$ \textbf{not} in the cache $C$}
			\STATE decide a block to overwrite with an algorithm like LRU; the victim block is written back to the output file.  
			\STATE $new\_block\leftarrow$ $request\_oram\_block(block\_id)$
			\STATE add $new\_block$ to the cache
			\ENDIF
			\STATE $block\_reference$ $\leftarrow$ find the $block\_id$ in the cache
			\STATE \textbf{return}  $block\_reference$
		\end{algorithmic}
		\hrule
		\begin{algorithmic}[1]
			\STATE \textbf{Function} $\mathrm{AddRecordToBlock} (record)$  
			\IF{working block $B$ is \textbf{full}}
			\STATE evict LRU and write out the victim block
			\STATE copy $B$ to the cache
			\STATE clear the working block 
			\ENDIF
			\STATE add $record$ to the working block
			\STATE \textbf{return} $working\_block\_id$
		\end{algorithmic}
		
	\end{algorithm}
	
	\begin{algorithm}
		\caption{HashMap-based KMeans in Enclave}\label{alg:oram_cache}
		\begin{algorithmic}[1]
			\STATE \textbf{Function} $\mathrm{KMeans} (centroid\_file, coordinates\_file)$
			\STATE initialize $ORAM$ and $Cache$ block
			\STATE load initial $centroids$ from $centroid\_file$
			\FOR{all $block$ in $coordinates\_file$}
			\STATE $points$ $\leftarrow$ ParseCoordinates($block$)
			\FOR{each $pt$ in $points$}
			\STATE $centroid\_index$ $\leftarrow$ FindNearestCentroid($pt$, $centroids$)
			\STATE LocalMap[ $centroid\_index$ ] $\leftarrow$ $pt$
			\ENDFOR
			\STATE $combined\_points$ $\leftarrow$ aggregate points under same centroid
			
			\FOR{ each ($centroid\_index$, $point$) in $combined\_points$ }
			\IF{ $centroid\_index$ not \textbf{not} in $HashMap$}
			\STATE $record$ $\leftarrow$ ($centroid\_index$, $point$)
			\STATE $id$ $\leftarrow$ AddRecordToBlock($centroid\_index$)
			\STATE $HashMap[$centroid\_index$]$ $\leftarrow$ $id$
			\ELSE
			\STATE $id$ $\leftarrow$ $HashMap[$centroid\_index$]$
			\STATE $block\_reference$ $\leftarrow$ GetBlock($id$)
			\STATE update block in $block\_reference$ with ($centroid\_index$, $point$)
			\STATE Write($block$)
			\ENDIF
			\ENDFOR
			\ENDFOR
			\STATE write all blocks from $ORAM$ to $centroid\_file$
			
		\end{algorithmic}
		
	\end{algorithm}

\end{document}